\documentstyle[12pt]{article}
\newcommand{\be}{\begin{equation}}
\newcommand{\ee}{\end{equation}}
\newcommand{\bea}{\begin{eqnarray}}
\newcommand{\eea}{\end{eqnarray}}
\newcommand{\ba}{\begin{array}{l}}
\newcommand{\ea}{\end{array}}
\newcommand{\bb}{\begin{thebibliography}}
\newcommand{\eb}{\end{thebibliography}}
\newcommand{\ci}[1]{\cite{#1}}
\newcommand{\lab}[1]{\label{#1}}
\newcommand{\re}[1]{(\ref{#1})}

\newcommand{\Ds}{\displaystyle}

\def \ins#1#2#3#4#5#6 {
  \begin{figure}[#5]
   \begin{center}
    \begin{minipage}{#3}
     \hspace{3.5cm} \vbox to #4{
     \vskip -1cm
     \special{em:graph #2 } } \\
          \caption{#6}
            \vspace{-0.3cm}
              \end{minipage} \end{center}
          \label{f}
                \end{figure}}

\begin{document}
\title{A ROTATED SKYRMION AND  PIONS-NUCLEON INTERACTIONS \\
}
\author{
 M.M.MUSAKHANOV\thanks{Permanent address: Theor.Physics Dept.,
Tashkent State University, Tashkent 700174, Uzbekistan,
e-mail: yousuf@iaph.silk.org}\\
Abdus Salam ICTP, Strada Costiera 11, 34014 Trieste, Italy\\
email: yousufmm@ictp.trieste.it}
\maketitle
\begin{abstract}
It was shown by careful  consideration of
 the time--dependent classic solution for the rotated skyrmion, that such
solution at large distances has radiational component similar to the
Lienart-Viehart retarded potential in Electrodynamics. We
consider such solution as stationary phase configuration of the
pion field  in path integrals for the correlators of $n$
isovector axial currents $J_{\mu}^{A,i}$  and two nucleon
currents $J_N$. $n$ pion -- nucleon amplitudes are extracted
from these correlators. The careful account of the asymptotics
of this  stationary phase configuration of the pion field
(rotated skyrmion) in path integral leads to nonzero contribution
of this classical part of the total pion field  to pion-nucleon
amplitudes.  In result, this approach correctly reproduce pions
-- nucleon amplitudes(with the Born -- diagrams contributions,
too).  \end{abstract}

\section{Introduction}
    The skyrmion or skyrmion-like models of a nucleon has a many attractive
features and was discussed very much (see review \ci{mus}, for example). 
Among them the most powerful are chiral quark models, inspired by QCD.

      We will consider pions-nucleon interactions in the framework of
   the $SU(2)$  -- effective chiral QCD model  \ci{cQCD}.  In this model
   the radius of SBCI is much smaller than the
 radius of confinement   $R_{conf}$.  It was shown that
 the interaction of valence
 quarks with pion field in nucleon confines them  at the
 distances much smaller than   $R_{conf}$ (chiral confinement) \ci{cQCD}.
 The color confinement   forces can be accounted as a
 perturbation  in this case.

 The Lagrangian of the effective chiral QCD  has the form of the nonlinear
 quark $\sigma$  -- model describing interacting quark $\psi$ and isovector
 pion $\vec \phi$ fields without kinetic term for the pion field:
\be L=\bar\psi(i \gamma \cdot \partial +
 M_q \exp(i \gamma_5 \vec\tau \vec\phi )) \psi .
\lab{L}
\ee
  The quantity $M_q$ in fact is a function of the   momentum
 $p$. This function $M_q (p)$ in the Euclid space
 can be approximated by $\Theta $-- function
 and contains two main parameters:
 the effective quark mass $M_q(0)$ and a cutoff $p_0$, which is defined
 from the condition $M_q(p_0)=0$.
Integration over quark fields in the partition function generates, as
 usual, the effective action
 $$S_{eff}[\phi |B=0] = \Ds \int dx L_{p}$$
 for the states with baryon number $B=0$, where Lagrangian
\be
L_{p} =
 -3\ln\det[i \gamma \cdot \partial +
 M_q \exp(i \gamma_5 \vec\tau \vec\phi) ] =
 { 1 \over 4} f^2 \, Tr \, L_\mu L^\mu  + ...
\lab{L_p}
\ee
 Here $f$ is the charge pion decay constant,  $L_\mu = U^+ \partial_\mu U$  
and $SU(2)$  -- valued matrix $U$
  can be represented as
 $U=e^{i \vec \tau \vec \phi }$.

 The $M_{q}(0)=340Mev$ and $p_{0}=600Mev$ were fixed from the well known
 quantities $f=93Mev$ and the electric charge radius of pion
 $r_{\pi}=(300 Mev)^{-1}$.

 The calculation of the nucleon properties is related with the calculation
 of the correlator of the 3 -- quarks currents with the quantum numbers of the
 nucleons $J_N$.
 Integration over quarks in this correlator gives
\be
\Pi_N(x,y) = \Ds \langle J_N (x) J_N^+ (y) \rangle
         = \Ds \int D \vec \phi \prod \limits_{i=1}^{3}G_{i}(x,y |\phi)
e^{iS_{eff}[\phi |B=0]}
\lab{Pi_N}
\ee
 Here $G_{i}(x,y |\phi)$ is the quark propagator in the external field 
$\phi$ and
 $S_{eff}[\phi |B=0]$ -- was defined in the  Eq. \re{L_p}).
 So, the effective action for the states with baryon number $B=1$
\be
S_{eff}[\phi , x, y |B=1] =
 S_{eff}[\phi |B=0] -iln \prod \limits_{i=1}^{3}G_{i}(x,y |\phi).
\lab{effaction}
\ee
 For a stationary external pion field and in the Euclid space 
 ($\vec x = \vec y = 0,$ $x_4 = T \to \infty ,$ $y_4 = 0$)
\be
G_{i}^{E} \sim \exp(-E_{val}[\phi] \cdot T),
S_{eff, E}[\phi |B=0] =   E_{vac}[\phi] \cdot T.
\lab{G_i}
\ee
 The total energy of such a state consists
 of contributions from the 3 valence quarks ($3E_{val}$) and polarization
 of quark vacuum ($E_{vac}$).  Both of them are functionals of the external
 pion field.
 Integration in the saddle--point approximation over pion field can be
 reduced to the minimization of the total energy over some suitable trial
 fields $\phi$ :
\be
M_N ^0 = \mbox{min}_{\phi} (3 E_{val} [\phi] + E_{vac} [\phi])
\lab{minim}
\ee
 The Skyrme model teaches us to choose this field in the form of the
chiral soliton (skyrmion):
\be
\ba
U_{0}(\vec x)=\exp i\vec\tau\vec n\phi _{0} , \,\,
\vec n = {\vec r
/{r}}, \\
\phi_{0}(0) = \pi ,
 \phi_{0}(r)\sim { - 3g_A
/{(8\pi f^2 r^2 )}} , (r \to \infty ).
 \ea
 \lab{phi_0}
 \ee
 Asymptotic of $\phi_{0}$ is in fact the Goldberger-Treiman relation,
  where $g_A$ is
 the axial charge of the nucleon.
This soliton field configuration has unit topological number
  \ci{Skyrme1}.
 The simplest trial function is  \ci{cQCD}:
\be
\phi_{0} = 2 \arctan ({r_0\over r})^2
\lab{F}
\ee

In the Effective chiral QCD model valence quarks
in the nucleon are confined in the region of order $0.5 fm$
due to interaction effects with their own pion field (chiral confinement).
The properties of the nucleon in this case resemble chiral soliton --
skyrmion very much.

It is useful to compare this and Skyrme models.
  The Lagrangian of Skyrme model has the form:
\be
L_{sk} = { 1 \over 4} f^2 \, Tr \, L_\mu L^\mu  +
         {1 \over 32 e^2} Tr [ L_\mu, L_\nu]^2 ,
\lab{L_sk}
\ee
   (Note, that the second term in \re{L_sk} is
 needed for the stability of the soliton).

 The energy of the trial field configuration  \re{F} is
$E= A\, f^2\, r_{0} + {B
/{(e^2 r_0})}$.
 The first term just coincides with the $E_{vac}$, and the second one
 simulates $E_{val}$.

 The main difference between
 Skyrme and this models is that Skyrme model action is the
 same for the pion and
 nucleon sectors. In contrast, the action of the  effective chiral QCD model
 has different forms
 for these sectors because the  nucleon sector has an additional
 contribution of valence quarks.

 \section {Time-dependent classic solution for the rotated
 skyrmion}

 We will consider classic time -- dependent solution
 of the chiral fields equations of $S_{eff}[\phi |B=1] $.
 We suppose that static equations
 of motion has topologicaly nontrivial solution -- skyrmion
 $U_{0}(\vec x)$.  We suppose that the time dependence for the rotated 
skyrmion at the
 small distances $x \sim r_{0}$ ($ r_{0}$ was defined at Eq. \re{F}) 
can be introduced into
 the static solution $U_{0}(\vec x)$ accordingly Ansatz:
 $$U(\vec x,t) = A(t)U_{0}(\vec x)A^{+}(t).$$
 $SU(2)$ matrix  $A(t)$ has a
 meaning of the collective coordinate of the rotation of the skyrmion
 as a whole in the isospace(rigid rotation) with the frequency
$\vec \Omega = -iSpA(t)\partial _{0}A^{+}(t)\vec\tau.$ We suppose that 
$\Omega r_{0} << 1$ (slow rotation).

 It is clear that
 the  solution of  the chiral field equations in the static case is
 equivalent to the minimization of the classic mass $M_s$ over
 parameter $r_{0}$ of the suitable skyrmion-like trial function the Eqs.
\re{phi_0}, \re{F}.
 The parameter
 $r_{0}$ has clear meaning of the size of the  skyrmion.
 Shall we consider the $$U(\vec x,t) =\exp
 i\vec\tau\vec\phi(\vec x, t) = \exp i\tau_{i}R_{ij}(t) n_{j}
 \phi_{0}(x) $$ at large distances in more detail.
 At large distances $\Omega ^{-1}>>x>>r$ we can neglect by nonlinearity
 of the pion fields and
 \be
 \phi_{i}(\vec x, t) = R_{ij}(t)n_{j}
 (r/x)^2
 \lab{asympt} \ee

 On other hand  the equation for the
 $\vec\phi(\vec x, t)$ at large distances has the form of the
 usual wave equation:
 \be
 (\Delta  - \partial^{2}/\partial t^{2})\phi_{i}(\vec x, t) = 0
 \ee
 It is clear that we can
 improve the asymptotic  \re{asympt} for the distances $x\sim 
\Omega^{-1}$ to take it
 in the form similar to the retarded  Lienart-Viehart
 potentials:
 \be
 \phi_{i}(\vec x, t) = - \partial_{j} R_{ij}(t-x)(r/x)
 \ee
 As usual the account of retarded time ( which is the time of the
 propagation of field from source to the point $\vec x$)
 leads to the $1/x$ dependence
 of fields $\phi _{i}(\vec x, t)$ at large distances. This
 component of fields can be named as radiating one.

 From the large distances rotated skyrmion can be considered as
 rotated dipole.

 In general case time--dependent solution must have radiating
 component and such skyrmion have to lost his energy by
 emitting of the pions.
 The reason to have stable solution for the nucleon is the same as in
 atom physics. It is well known that nucleon will be the lowest fermion
 state of the quantum rotating skyrmion.
 So, we have the time dependent solution $\phi_{i}(\vec x,t)$in the form:
 \be
 \ba\Ds
 {\rm at} \,\, 0<x<r \,\, (r_{0} <<r<<\Omega ^{-1}) \\ \\
\Ds
 \phi_{i}(\vec x,t) \, =\, R_{ij}(t)\,n_{j}\,2 \arctan
\frac{r_{0}}{2\, x^2} \\ \\
 {\rm and} \,\, {\rm at} \,\,  x>>r \\ \\ \Ds
 \phi_{i}(\vec x, t) \, =\, -\, \partial_{j}\, R_{ij}(t-x)\, \frac{r}{x}
 \ea
 \lab{solution}
 \ee
 Let us calculate the action by using the Eq. \re{solution}:

 \be
 \ba
 S = \Ds \int \limits_{t_{in}}^{t_{f}}dt\Ds \int \limits_{x<r} d^{3}x
 L(\phi_{i}(\vec x. t) = R_{ij}(t) n_{j}\phi_{0}(x)) +
  f^{2}/2\Ds \int \limits_{x>r} d^{3}x
 \partial_{\mu}\phi_{i}(\vec x. t)\partial_{\mu}\phi^{*}_{i}(\vec x. t) = \\
 \Ds \int \limits_{t_{in}}^{t_{f}}dt[(I(\vec \Omega)^{2}/2 - M_{s})_r +
 f^{2}/2\Ds \int \limits_{x>r} d^{3}x \partial_{\mu}(\phi_{i}(\vec x. t)
 \partial_{\mu}\phi^{*}_{i}(\vec x. t))
 \ea
 \ee
 We calculated the former integral by accounting of the Eq. of motion at
 large distances and by the integration by part.
 For this part of the action we have
 \be
 \ba
 f^{2}/2\Ds\int \limits_{t_{i}}^{t_{f}}dt
 \int \limits_{x>r} d^{3}x \partial_{\mu}(\phi_{i}(\vec x. t)
 \partial_{\mu}\phi^{*}_{i}(\vec x. t))| = \\
 f^{2}/2(\Ds \int \limits_{x>r} d^{3}x(\phi_{i}(\vec x. t)
 \partial_{0}\phi^{*}_{i}(\vec x. t))|_{t_f} -
 \int  \limits_{t_{i}}^{t_{f}}dt\oint_{(x=r)}
 d \vec n x^{2} (\phi_{i}(\vec x. t)
 \partial_{x}\phi^{*}_{i}(\vec x. t)) ) = O(1/r )
 \ea
 \ee
 The terms of order $O(1/r)$ reduce the dependence of first term
 in the action $S$ on $r$.

 It is easy to calculate the energy emitted by rotated
skyrmion in the unit of time by calculation of the energy-momentum tensor 
$$T_{\mu\nu} = f^{2}/2(Tr\partial_{\mu}U\partial_{\nu}U^{+} -
1/2 g_{\mu\nu} Tr\partial_{\alpha}U\partial_{\alpha}U^{+} ) +
O(\partial^{4}).$$ 
At $x>>r_{0}$ \\
$T_{00} = f^{2}/2(\partial_{0}\phi_{i}\partial_{0}\phi_{i}^{+} +
\partial_{j}\phi_{i}\partial_{j}\phi_{i}^{+}) $,
$T_{0j }= f^{2}\partial_{0}\phi_{i}\partial_{j}\phi_{i}^{+} $\\
$\oint_{S(x=r)}dS_{m} T_{0m} = - \alpha f^{2}
( 2 \vec \Omega^{4} (t-r)+ 2 (\partial_{0}\vec \Omega)^{2}(t-r) +
3\partial_{0}(\vec \Omega(t-r))^{2})$\\

 In fact we need components $T_{0n}$ at large distances to
calculate $dE/dt = \oint_{S(x=r)} dS_{n}T_{0n}$. \\

Energy at the distances $x>r$:\\
$\Delta E(t) =\Ds \int \limits_{x>r} dx \alpha f^{2}(2 \vec \Omega^{4} (t-r)+
2 (\partial_{0}\vec \Omega (t-r))^{2} + 3 
\partial_{0}(\vec\Omega(t-r))^{2})$\\ it is nonzero if $t > r$ \\
(Another formula for the calculation of the same quantity is \\
 $\Delta E(t) = \Ds \int  \limits_{r}^{t}dt' r^{2}d\vec n n_{m}T_{0m }$  \\
and give the same answer).
We can neglect by it because $ \Delta E(t) \sim O( \Omega^{4} ) $ and we 
suppose that $\partial_{0}\Omega << \Omega^{2}$, $\Omega r_{0} << 1$. \\ 
With this accuracy 
we can consider classic rotated skyrmion as stable stationary state.
In this case the quantization of this object is well defined procedure.
We will follow to the method of Ref. \ci{rajaraman}, where was considered the
similar problem but with $U(1)$ internal symmetry. The straightforward
generalization of this results means in our case that we must to
minimize over variational parameter $r_{0}$ the expression for
the total energy of rotated skyrmion with account of the kinetic energy
of the rotation of the skyrmion and the contributions of the small quantum
transverse (to zero-mode $A$) fluctuations. In this case the parameter
$r_{0}$ has meaning of the size of a rotated skyrmion.
\be
\ba
min E_{tot}[r_{0}] = M_{s}[r_{0}] + \hat J^{2}/2I[r_{0}] + \\
(contributions \, of \, the \, quantum \, transverse \, fluctuations)
\ea
\ee
 Quantization of the rotational motion over $A(t)$ leads to the
spectrum of
 the quantum rotator $$E = M_s + T(T+1)/2I$$
 with $T=S$ ($S$ is a spin, $T$--isospin and $I$ -- classical
 part of the moment of inertia).
 Wave functions are the Wigner $\cal D$-- functions:
\be
W.F. = {\cal D}^{S=T}_{s_3,t_3}(A).
\lab{WF}
\ee
 The nucleon is identified with the lowest fermion state with $S=T={1\over 
2}$.

\section
{ Pions -- nucleon interactions in the Effective Chiral QCD model}

From the general point of view any $n$ pions -- nucleon amplitude can be
extracted from the appropriate correlator of $n$ isovector axial currents
$J_{\mu}^{A,i}$
(or their divergencies) and two nucleon currents $J_N$   \ci{scat3}
\be
\ba
A_{\mu_1 ...\mu_n}(p_1 , p_2 , q_1 ,... , q_n )
= \int d^4 y_{1}d^4 y_{2} d^4 x_{1}...d^4 x_{n}
\exp i(q_1 x_1 + ...+q_n x_n - p_1 y_1 + p_2 y_2 ) \\ \\
\Ds \langle TJ_N ^+ (y_1 ) J_N (y_2 )
J_{\mu_1}^{A,i}(x_1 )
...J_{\mu_n}^{A,j}(x_n )\rangle
\ea
\lab{A_mu}
\ee
(In the calculations of this part of the section we will use Minkowsky space
formulation).
The $n$ pions -- nucleon amplitude $A_n (p,q)$ is a residue in the mass shell
poles of the correlator \re{A_mu}.
For the calculation of this correlator we introduce external axial-vector
field $A_{\mu}={\tau_{i}
/2}A^{i} _{\mu} (x)$. The interaction of quarks
with this field can be
introduced by substitution:
$\partial_{\mu} =>\partial_{\mu} - i\gamma_{5}A_{\mu}$
in the Lagrangian \re{L}.

In this case, we calculate  the correlator in Eq. \re{A_mu} by using the
formula:
\be
\ba
\langle TJ_N ^+ (y_1 ) J_N (y_2 )
J_{\mu_1}^{A,i_1}(x_1 )
...J_{\mu_n}^{A,i_n}(x_n )\rangle = \\
{\delta \over \delta A^{i_1} _{\mu_1} (x_1)}...
{\delta \over \delta A^{i_n} _{\mu_n} (x_n)}
\langle TJ_N ^+ (y_1 ) J_N (y_2 )\rangle _{A} |_{A=0} ,
\ea
\lab{delta}
\ee

where
$\langle TJ_N ^+ (y_1 ) J_N (y_2 )\rangle _{A}$ is
 the correlator of the 3 -- quarks currents with the quantum numbers of the
 nucleons $J_N$ in the presence of the external axial-vector field $A_{\mu}$.
 Integration over quarks in this correlator gives

\be
\langle TJ_N ^+ (y_1 ) J_N (y_2 )\rangle _{A}
= {{ \Ds \int D \vec \phi
e^{iS_{eff}[\phi , A , y_1 , y_2 |B=1]}} \over
\Ds \int D \vec \phi e^{iS_{eff}[\phi |B=0]}},
\lab{ext}
\ee

 where  effective action in presence of external axial field $A$
(see also Eq.\re{Pi_N})
 \be
 S_{eff}[\phi , A , y_1 , y_2 |B=1] =
 -i ln\prod \limits_{i=1}^{N_c}G_i(y_1 ,y_2 |\phi , A) +
 S_{eff}[\phi , A |B=0]
 \lab{B=1}
 \ee

 In order to have the effective action
 in the presence of the external axial-vector field $A$ we must replace in
 $S_{eff}[\phi , y_1 , y_2 |B=1]$
 $\partial _{\mu}U$ by $\partial _{\mu}U + A_{\mu}U + UA_{\mu}$ etc.
 \ci{pak}.

For the calculation of the correlator \re{delta} near pion's mass-shell
region it is enough
to account such a substitution only in the kinetic term in the Eq.\re{L_p}.
In Eq. \re{delta} we can replace ${\delta
/(\delta A^{i} _{\mu} (x))}$ by
$(-2if){\delta
/(\delta \partial_{\mu}\phi^{i} (x))}$.
It is clear, that in the result we have
\be
\ba
A_{\mu_1 ...\mu_n}(p_1 , p_2 , q_1 ,... , q_n )
= \int d^4 y_{1}d^4 y_{2} d^4 x_{1}...d^4 x_{n}
\exp i(q_1 x_1 + ...+q_n x_n - p_1 y_1 + p_2 y_2 ) \\
= (-2ifq_{\mu_1})...(-2ifq_{\mu_n}){\Ds \int D \vec \phi \phi_{i_1} (x_1 )...
\phi_{i_n} (x_n )e^{iS_{eff}[\phi , y_1 , y_2 |B=1]} \over
\Ds \int D \vec \phi e^{iS_{eff}[\phi |B=0]}},
\ea
\lab{pi}
\ee
A straightforward calculation of the Eq. \re{pi}
leads to the following formula for $A_n (p,q)$
\be
\ba
A_n (p,q)
= \int d^4 x_{1}...d^4 x_{n} d^4 y_{1}d^4 y_{2}
\exp i(q_1 x_1 + ...+ q_n x_n -p_1 y_1 + p_2 y_2 ) \\
q_1 ^2 ...q_n ^2 (M_N ^2 - p_1 ^2 )(M_N ^2 - p_2 ^2 )
{{\Ds \int D \vec \phi \phi_i (x_1 )... \phi_j (x_n )
e^{iS_{eff}[\phi , y_1 , y_2 |B=1]} \over
\Ds \int D \vec \phi
e^{iS_{eff}[\phi |B=0]}}},
\ea
\lab{A_n}
\ee
The stationary phase method is suitable for the calculations of Eq.\re{A_n}.
It is clear from previous section that in the numerator stationary phase
configuration of pion field is  Eq.\re{solution}.
In the denominator it is $\vec\phi = 0$. The measure in the path integral
in the numerator consist of from measure of the
collective translations $R(t)$ and rotations $A(t)$
and transverse to these modes
the quantum fluctuations $\vec\phi ^q$. Total field
$$\vec\phi (x)= \vec\phi ^s (x) + \vec\phi ^q (x),$$
$\phi^{s}$ is accounted the translations $R(t)$ too and finally has a form 
\be \ba
\Ds
\phi ^{s}_{i} (x)\, =\, 
 R_{ij}(t)\, n_{j}2\, \arctan \,\frac{r_{0}}{2(\rho (t))^2} \,\,\,\,\,\,
\,\,\,  {\rm at} \,\,\,\, 0 <\rho <r 
\\
 {\rm and}
\\\Ds
\phi ^{s}_{i} (x) \, =\,   -\, \partial_{j}\, R_{ij}(\tau)\,
\frac {r^{2}_{0}}{\rho (\tau)}, 
\,\,\,\,\,\,\,\,\, {\rm at} \,\,\,\,  \rho >>r .
\ea
\lab{phi}
\ee
Here $\rho (t) = x - R(t)$ and $\tau = t - \rho (\tau )$
is retarded time.

The expansion of the numerator of the Eq.\re{A_n} around $\vec\phi ^s$ over
$\vec\phi ^q$ leads to two main parts of amplitude $A_n (p,q)$. The first one
contains only $\vec\phi ^s$ and second one --
leading order over $ \vec\phi ^q.$ 
Accordingly it
the measure and the effective action in Eqs. \re{pi}, \re{A_n} can be 
written as $$D \phi = D \vec R(t) D A(t) D \phi ^q (\vec x, t),$$
$$S_{eff}[\phi , y_1 , y_2 |B=1] = S_s + S_q.$$
Here
$$S_s = \Ds \int \limits_{y_{1,0}}^{y_{2,0}}dt(-M_s + M_s \vec V^2 /2
+ I \vec \Omega ^2 /2), $$
$$S_q =
\Ds \int d^4 x tr \phi ^q (x) 1/2 \hat L [\phi_s ] \phi ^q (x) +
O(\phi _{q}^3 ),$$ 
where $M_s$ and $I$ are classical parts of the nucleon mass and
moment of inertia; $\vec V=d \vec R/dt$ and
$\vec \Omega $ are nucleon velocity and
nucleon angular velocity; $\hat L [\phi_s ]$ -- the quadratic form for
the quantum field $\phi ^q$.

Let us consider the simplest amplitude $A_1 (p,q)$:
\be
\ba
A_1 (p,q)
= \int d^4 x d^4 y_{1}d^4 y_{2}
\exp i(qx  -p_1 y_1 + p_2 y_2 ) \\
q_{1}^{2}(M_{N}^{2} - p_{1}^{2})(M_{N}^{2} - p_{2}^{2})
\Ds \int D \vec R(t) D A(t) D \phi^{q} (\vec x, t) \phi^{s}_{i} (x)
\exp i(S_{s} + S_{q})
\ea
\lab{A_1}
\ee

Integration over $\phi ^q$ leads to the quantum correction of the
nucleon mass. Integration over the nucleon position $\vec R(t)$ and
orientation $A(t)$ provide us with nucleon propagator:
$$
\ba
G(t_2 , R_2 =R(t_2 ), A_2 =A(t_2 ); t_1 , R_1 =R(t_1 ), A_1 =A(t_1 )) = \\ \\
\sum \limits_{T,T_3 } \langle R_2 , A_2  | T,T_3 \rangle \langle T,T_3 |
R_1 , A_1 \rangle \int d^3 \vec p /(2\pi )^3  \\ \\
\exp i(\vec p (\vec R_2 - \vec R_1 ) - (M_s + \vec p^2 /2M_s +
T(T+1)/2I )(t_2 - t_1 )
\ea
$$

The Eq. \re{A_1} can be written in the form:
\be
\ba\Ds
A_1 (p,q)
= \int d^4 x d^4 y_{1}d^4 y_{2}
\exp i(q x  -p_1 y_1 + p_2 y_2 )q_1 ^2 (M_N ^2 - p_1 ^2 )(M_N ^2 - p_2 ^2 ) \\
\Ds \,\,\,\,\,\,\, 
G(y_{2,0} , \vec y_2 , A_2 ; \tau (x_0) , \vec x, A ) \phi ^s (x)
G(\tau (x_0) , \vec x, A; y_{1,0} , \vec y_1 , A_1 )
\ea
\lab{A_11}
\ee

It is important feature of Eq. \re{phi} that we used the classical field
$\phi ^s (x)$ in the form of the retarded Lienard--Viechard potential.
By this reason in Eq. \re{A_11} retarded time
$ \tau = x_0 - | \vec x - \vec R (\tau )|$ is used.
We perform the integrations in this Eq. by natural introducing
of the new variables:
$\tau$ instead of $x_0$ and $\vec z=\vec x - \vec R(\tau )$.
Nucleon propagators cancel the factors $(M_N ^2 - p_{1,2}^2 )$,
integrations over $\vec x$ and $\tau$ provide energy-momentum
conservation $\delta$ -function. At last, the integral over $\vec z$ has
the form:
\be
\vec \tau \vec q  \Phi(q) =
\int d^3 \vec z \exp i(\vec q \vec z - q_{1,0} |\vec z|) \phi _0 (\vec z),
\lab{pole}
\ee

where $\phi _0 (\vec z )$ is given by Eq. \re{phi_0}.
It is clear that
in the limit $q^{2}\displaystyle\rightarrow\, 0$ it is sufficient
to save asymptotic of
$\phi_0$ Eq. \re{phi_0} only due to $q^2$ factor in Eq.\re{A_11}.
In this case we have an usual answer for the pion -- nucleon vertices
\be
\Gamma (A) = 3g_A /4f Tr( \tau_a A\tau_b A^+ q_b )
\ee
\lab{c8}
It is clear from Eq. \re{pole}
that the off -- mass shell (nonzero $q^2$)
pion -- nucleon form -- factor is given
by the formula
\be
F_{\pi NN} (q^{2}) = 2f /(3g_A) (q^2 \Phi (q^2 ))
\ee
\lab{c9}
Certainly, this form-factor is unphysical and depend on the definition
of the off-mass shell continuation, as it is clear from the previous 
discussion of the correlator, Eq. \re{pi} and the amplitudes, Eq. \re{A_n}. 
Now, it is rather easy to discuss pion-nucleon 
scattering. In the pre-exponent factor in the general Eq.\re{A_n} we have
the contribution from
$\phi ^s (x_1 )\phi ^s (x_2 )$
and from
$\phi ^q (x_1 )\phi ^q (x_2 )$
terms \\
(it is clear that mixed terms
$\phi ^q (x_1 )\phi ^s (x_2 )$
and
$\phi ^s (x_1 )\phi ^q (x_2 )$
contributions are zero).

We can repeat and extend of the above calculations of the
contributions of the $\phi ^s $ for two stage at
every point $x_1$ and $x_2$. Time ordering leads to two type of the
contributions -- $s$--channel and $u$--channel poles.
We would have naturally it in the form of the
Born pole-diagrams with sequence of the intermediate different rotated states
of nucleon ($S=T=1/2, 3/2,...$).

The calculation of the contribution of the
$\phi ^q (x_1 )\phi ^q (x_2 )$  term to the pion--nucleon
scattering amplitude is related with the calculation
of the inverse operator $\hat L^{-1} [\phi^s ]$
(see e.g. \ci{scat5}).
This one is
usual potential scattering problem with
$$\hat L [\phi^s ] =  \delta ^2 S_{eff}[\phi |B=1]/ \delta ^2 \phi |_{\phi ^s}$$.

Accordingly Ref.\ci{cQCD}
$S_{eff}[\phi |B=1]$ and $\hat L [\phi^s ]$ consequently
get the short ranged contribution from valence
quarks term ($<0.5fm$) and
long ranged one from vacuum polarization term ($>0.5fm$).

I would like to thank S. Bashinsky for the interesting discussion. 
This work was supported in part by the
ICTP, Grant INTAS-96-0597ext
and State Committee for Science and Technology of Uzbekistan.

\end{document}